\documentclass[fleqn,usenatbib]{mnras}

\usepackage{newtxtext,newtxmath}

\usepackage[T1]{fontenc}

\DeclareRobustCommand{\VAN}[3]{#2}
\let\VANthebibliography\thebibliography
\def\thebibliography{\DeclareRobustCommand{\VAN}[3]{##3}\VANthebibliography}

\usepackage{graphicx}	

\usepackage{amsmath,amsfonts}
\usepackage{setspace}
\usepackage{subfigure}
\usepackage[subfigure]{tocloft}
\usepackage{lineno}
\usepackage{multirow}

\usepackage[latin1]{inputenc}

\usepackage{units}

\title[Advances in control of a Pyramid SCAO system]{Advances in control of a Pyramid Single Conjugate Adaptive Optics system}

\author[Guido Agapito et al.]{
Guido Agapito,$^{1}$\thanks{E-mail: guido.agapito@inaf.it}
Fabio Rossi,$^{1}$
Cedric Plantet,$^{1}$
Alfio Puglisi,$^{1}$
Enrico Pinna$^{1}$
\\
$^{1}$Osservatorio Astrofisico di Arcetri (INAF), Largo E. Fermi 5, Firenze, Italy, 50125
}

\date{Accepted XXX. Received YYY; in original form ZZZ}

\pubyear{2021}

\begin{document}
\label{firstpage}
\pagerange{\pageref{firstpage}--\pageref{lastpage}}
\maketitle




\begin{abstract}

Adaptive optics systems are an essential technology for the modern astronomy for ground based telescopes. One of the most recent revolution in the field is the introduction of the pyramid wavefront sensor. The higher performance of this device is payed with increased complexity in the control. In this work we report about advances in the AO system control obtained with SOUL at the Large Binocular Telescope. The first is an improved Tip/Tilt temporal control able to recover the nominal correction even in presence of high temporal frequency resonances. The second one is a modal gain optimization that has been successfully tested on sky for the first time. Pyramid wavefront sensors are the key technology for the first light AO systems of all ELTs and the reported advances can be relevant contributions for such systems.     


\end{abstract}







\section{Introduction}\label{intro}

Modern Single Conjugate Adaptive Optics (SCAO) systems are a category of Adaptive Optics (AO) systems designed to get diffraction-limited resolution on near infrared wavelengths.
These kind of systems are equipped with hardware and software designed to provide such performance: they comprise high actuator density, fast Deformable Mirrors (DMs) and Natural Guide Star (NGS) WafeFront Sensors (WFSs) with high spatial and temporal sampling to reduce fitting, temporal and aliasing errors.
In 2010, FLAO (\cite{2011SPIE.8149E..02E}), the first one equipped with a Pyramid WafeFront Sensor (PWFS \cite{Ragazzoni96}), delivered unmatched high contrast images in H band, opening a new era in the field (\cite{doi:10.1146/annurev-astro-081811-125447}).
Then, many others adopting the PWFS and succesfully brought it on sky (magAO \cite{2012SPIE.8447E..0XC, 2013aoel.confE..91C}, SCExAO \cite{10.1117/12.2314282, 10.1117/12.2529689, Jovanovic_2015}, Keck \cite{2020JATIS...6c9003B}, MagAO-X \cite{10.1117/12.2312992}).
Nowadays, the most commonly used WFS for SCAO systems with NGS is the PWFS.
This includes future SCAO systems in development for the 8-m class telescope (GPI 2.0 \cite{10.1117/12.2562578}, SPHERE+ \cite{boccaletti2020sphere}) and all those for the next generation of Extremely Large Telescopes (ELT-MICADO \cite{2021Msngr.182...17D}, ELT-HARMONI \cite{2021Msngr.182....7T}, ELT-METIS \cite{2021Msngr.182...22B}, TMT \cite{2018SPIE10703E..3VC}, GMT-NGSAO \cite{2014SPIE.9148E..2MP}).
In this work we report about the advances in AO control developed and tested on the SOUL (Single conjugate adaptive Optics Upgrade for LBT \cite{10.1117/12.2234444}) system. This system is the upgrade of the 4 First Light Adaptive Optics (FLAO) systems of the Large Binocular Telescope (LBT \cite{2010ApOpt..49..115H}) and it is described in  \cite{10.1117/12.2234444}.
The main new hardware features reduced RON and improved spatial and temporal sampling of the FLAO PWFS to get better performance on the whole magnitude range of the system (R<18).

A SCAO system performance is not only dependent on its hardware, but also on its control system.
It must reject turbulence, static aberrations and wind shake/vibrations measuring residual wavefront and reconstructing the incoming aberration.
This problem is well known, and all SCAO systems implement algorithms to optimize their control strategy.
In particular now the issue of optimizing the control of PWFS is very popular in the AO community because it has been recognized to be of high importance to optimize the image quality delivered by AO systems.
In brief, the PWFS provides higher sensitivity with respect to other sensors (\cite{Ragazzoni99}), however this sensitivity changes as a function of the wavefront correction (\cite{2001A&A...369L...9E}). This increases the complexity of the AO control and it has been the subject of many
recent works: \cite{2019SPIE11117E..0WJ,2020JATIS...6a0901S,2020arXiv200307228S,10.1093/mnras/staa843,10.1117/12.2563136,Deo2021} . 
The AO control optimization must minimize the propagation of all error terms in time-varying conditions (atmosphere, wind, vibrations, ...), adapting the free parameters.
SOUL is no exception.

The work is organized as follows.
In the first part of Sec. \ref{sec:controlIntro} we present the SOUL control system and the general strategy for setting up its free parameters.
The rest of the section focuses on the temporal control describing its structure and its implementation in the RTC (Sec. \ref{sec:tempControl}), the new solution developed for dealing with the peculiarities of the Tip/Tilt disturbances (Sec. \ref{sec:TTcontrol}) and our work on modal gain optimization in case of non-linear WFSs (Sec. \ref{sec:MGM}).
Then, Sec. \ref{sec:results} reports the results obtained with SOUL at LBT both with calibration source and on sky with NGS. Here we show the first modal gain optimization obtained on sky with the PWFS.

\section{Control system}\label{sec:control}

\subsection{The SOUL control system}\label{sec:controlIntro}
We consider the control system to be the supervisor that configures the AO system, closes the feedback loop and applies the temporal filtering.
The SOUL control ---
a modal \cite{RoddierBook1999} control with 672 degrees of freedom, that are the actuators of the Adaptive Secondary Mirror (ASM), and with 2400 inputs that are the x and y slopes coming from 1200 sub-apertures (approximately $40^2\frac{\pi}{4}$, where 40 is the number of sub-apertures on the diameter)
--- is configured trying to balance performance and robustness and
its configuration is made of a set of free parameters:
\begin{itemize}
    \item the integration time of the detector, which is the main way to regulate the amount of light received and hence the signal-to-noise ratio. It has an impact on temporal error and noise error.
    \item temporal control parameters, like
    gain, zeros and poles (see Sec. \ref{sec:tempControl}), that change the transfer function and has an impact on temporal, noise, aliasing errors.
    \item number of modes to be controlled (modal control).
    The number of controlled modes is reduced when the combination of noise and aliasing errors is larger than the additional fitting error.
    It can be achieved with zero-gain temporal control and non-controlled modes.
    \item detector binning mode/spatial sampling, which has an impact on noise, aliasing and fitting errors.
    \item circular modulation amplitude, which has an impact on sensitivity and linear range.
\end{itemize}
The last parameter is specific to systems equipped with a PWFS.

Our strategy for the choice of these parameters is: 
register the amount of detected flux at the beginning of an observation, and, based on this value, choose the configuration.
This strategy is required to cope with the features of the SOUL RTC: it is not able to change these parameters during closed loop operation, except for the gains of the temporal control.
On the other hand, the detected flux, that is strictly related to the signal-to-noise ratio, is the only parameters that can be quickly assessed in open loop.
Note that while the DIMM seeing measurements are available at the LBT,
turbulence conditions can drastically change during an observation,
so this information is not reliable enough to tune the system configuration once and for all at the beginning of the observation.
So all parameters except loop gains, that can be updated live as a function of the current status, are set up by a look-up table (see Tab. \ref{Tab:params}) and they are chosen to work properly in different conditions.
Poles and zeros of the temporal control make no exception and they are tabulated as a function of the integration time of the detector\footnote{an example is shown in Tab. \ref{Tab:roots}.} (which in turn depends on the detected flux).

In this work we focus on the temporal control, while the optimization of the other parameters, shown in Tab. \ref{Tab:params}, has already been exhaustively analysed in previous works for FLAO and SOUL systems 
~\cite{QuirosPacheco2010}, \cite{Esposito2010}, \cite{10.1117/1.JATIS.5.4.049001}, \cite{Pinna2019} and \cite{Agapito2019}. 

\begin{table*}
\caption{Summary of SOUL parameters as a function of natural guide star R magnitude.}
\label{Tab:params}
\begin{center}
\begin{small}
	\begin{tabular}{|l|c|c|c|c|c|}
		\hline
		\textbf{R} & \textbf{binning} & \textbf{no.}  & \textbf{no.} & \textbf{frequency} & \textbf{modulation} \\
		\textbf{magnitude} & \textbf{mode} & \textbf{sub-aps} & \textbf{modes} & \textbf{[Hz]} & \textbf{radius [$\lambda$/D]} \\
		\hline
        R$\leq$10.5 & 1 & 40 & 500 & 1700 & 3 \\
		\hline
        10.5<R$\leq$12.7 & 1 & 40 & 500 & 1700<f$\leq$500 & 3 \\
		\hline
        12.7<R$\leq$14.5 & 2 & 20 & 250 & 1250<f$\leq$750 & 3 \\
		\hline
        14.5<R$\leq$16.5 & 3 & 13 & 90 & 1000<f$\leq$250 & 3 \\
		\hline
        16.5<R$\leq$18.5 & 4 & 10 & 54 & 650<f$\leq$100 & 3 \\
		\hline
	\end{tabular}
\end{small}
\end{center}
\end{table*}

\subsection{Temporal control}\label{sec:tempControl}

SOUL control follows a modal approach as the one described in \cite{QuirosPacheco2010} for FLAO:
the reconstruction matrix (also known as control matrix) projects the measurement vector on a modal basis and, on this basis, the temporal control is applied independently on each mode.
Then, the outputs of the temporal filtering are projected on the actuators space by the so-called modes-to-commands matrix.
Hence, the temporal control in SOUL is made of a set of Single-Input Single-Output (SISO) controls.
Each element of this set is an Infinite Impulse Response (IIR) filter and its state-space representation with a null feedforward matrix is:
\begin{equation} \label{eq:stateIIR}
    \left. \begin{array}{l}
    \boldsymbol{x}(\kappa) = \boldsymbol{A} \, \boldsymbol{x}(\kappa-1) + \boldsymbol{B} \, \boldsymbol{u}(\kappa)\\
    \boldsymbol{y}(\kappa) = \boldsymbol{C} \, \boldsymbol{x}(\kappa)
    \end{array} \right\}
\end{equation}
where $\boldsymbol{u} \in \mathbb{R}$ is the filter input (in our case this is the modal measurement), $\boldsymbol{y} \in \mathbb{R}$ is the filter output (in our case this is the modal command), $\boldsymbol{x} \in \mathbb{R}^{N_{\mathrm{f}}}$ is the state vector (in our case $N_{\mathrm{f}}/2$ past values of $\boldsymbol{u}$ and $\boldsymbol{y}$), $\kappa$ is the temporal step, $\boldsymbol{A} \in \mathbb{R}^{N_{\mathrm{f}} \times N_{\mathrm{f}}}$ is the state update matrix, $\boldsymbol{B} \in \mathbb{R}^{N_{\mathrm{f}} \times 1}$ is the input matrix, $\boldsymbol{C} \in \mathbb{R}^{1 \times N_{\mathrm{f}}}$ is the output matrix and $N_{\mathrm{f}}$ is the filter state dimension.
The transfer function associated to the IIR filter is:
\begin{equation} \label{eq:tfIIR}
    \boldsymbol{H}\left(z\right)=\boldsymbol{C}\left(\boldsymbol{I}-\boldsymbol{A}z^{-1}\right)^{-1}\boldsymbol{B} = g\frac{\prod_{k=1}^{N_{\mathrm{\beta}}} (z-\beta_{\mathrm{k}})}{\prod_{l=1}^{N_{\mathrm{\alpha}}} (z-\alpha_{\mathrm{l}})}
\end{equation}
where $z$ is the variable of the Z-transform ($z^{-1}$ is equivalent to a time delay of one frame, $\boldsymbol{x}(\kappa) z^{-1} = \boldsymbol{x}(\kappa-1)$), $g$ is a scalar value known as filter gain, $\beta_k$ are the $N_{\mathrm{\beta}}$ roots of the numerator, known as zeros, $\alpha_{\mathrm{l}}$ are the $N_{\mathrm{\alpha}}$ roots of the denominator, known as poles and $N_{\mathrm{f}}=N_{\mathrm{\beta}}+N_{\mathrm{\alpha}}$  (we refer the interested readers to \cite{Goodwin2001}).

A particular and interesting case is when $N_{\mathrm{f}}=1$ (that means $N_{\mathrm{\beta}}=0$ and $N_{\mathrm{\alpha}}=1$).
In this case we get the so called leaky integrator and all the matrices become scalar values: $\boldsymbol{A}=\alpha$, $\boldsymbol{B}=g$ and $\boldsymbol{C}=1$. 
The state-space representation with a null feedforward matrix for a set of $N_{\mathrm{I}}$ leaky integrators is:
\begin{equation} \label{eq:stateINT}
    \left. \begin{array}{l}
    \boldsymbol{x}'(\kappa) = \boldsymbol{F} \, \boldsymbol{x}'(\kappa-1) + \boldsymbol{G} \, \boldsymbol{u}'(\kappa)\\
    \boldsymbol{y}'(\kappa) = \boldsymbol{I}_{N_\mathrm{I}} \, \boldsymbol{x}'(\kappa)
    \end{array} \right\}
\end{equation}
where $\boldsymbol{x}' \in \mathbb{R}^{N_\mathrm{I}}$ is the state vector, $\boldsymbol{u}' \in \mathbb{R}^{N_I}$ 
is the input vector, $\boldsymbol{y}' \in \mathbb{R}^{N_\mathrm{I}}$ is the output 
vector, $\boldsymbol{F}$ and $\boldsymbol{G}$ are diagonal matrices of respectively of the filter poles, $\boldsymbol{A}=\alpha$ (known in this case as forgetting factors), and of the filter gains, $\boldsymbol{B}=g$, and $\boldsymbol{I}_{N_\mathrm{I}} \in \mathbb{R}^{{N_\mathrm{I}} \times {N_\mathrm{I}}}$ is an identity matrix, because $\boldsymbol{C}=1$.

The implementation of the temporal control in the RTC is basically the same as FLAO, beside the ability to support the spatial and temporal sampling increase.
It is based on a few MVMs and relies on the following structure: 
\begin{equation} \label{eq:stateRTC}
    \left. \begin{array}{l}
    \boldsymbol{\zeta}(\kappa) = \boldsymbol{\Lambda} \, \boldsymbol{\zeta}(\kappa-1) + \boldsymbol{\Delta} \, \boldsymbol{\Xi} \, \boldsymbol{s}(\kappa)\\
    \boldsymbol{c}(\kappa) = \boldsymbol{\Sigma} \, \boldsymbol{\zeta}(\kappa)
    \end{array} \right\}
\end{equation}
where $\boldsymbol{\zeta} \in \mathbb{R}^{672}$ is the state vector, $\boldsymbol{s} \in \mathbb{R}^{2400}$ 
is the measurement vector and $\boldsymbol{c} \in \mathbb{R}^{672}$ is the command 
vector. 
The real time computer matrices are as follows:
\begin{itemize}
\item $\boldsymbol{\Lambda} \in \mathbb{R}^{672 \times 672}$ (state update matrix);
\item $\boldsymbol{\Xi} \in \mathbb{R}^{672 \times 2400}$ (designed to be the so-called reconstruction matrix $\mathcal{R}$);
\item $\boldsymbol{\Delta} \in \mathbb{R}^{672 \times 672}$ (gain matrix) is a diagonal matrix;
\item $\boldsymbol{\Sigma} \in \mathbb{R}^{672 \times 672}$ (designed to be the so-called modes-to-commands matrix $\mathcal{M}$).
\end{itemize}
As we stated before, these matrices, except for $\boldsymbol{\Delta}$, cannot be updated during an observation, as for the modulation amplitude, integration time and detector binning mode (they can be changed only in open loop).
So control filter state update and gain matrices must be chosen once per observation as a function of the detected flux (like the other free parameters).
Despite the fact that this implementation was designed to provide 672 filters with $N_{\mathrm{f}}=1$
(Eq. \ref{eq:stateRTC} is equivalent to Eq. \ref{eq:stateINT} with the additional reconstruction and modes-to-commands matrices), it is possible to implement filters with $N_{\mathrm{f}}>1$, but since the dimension of the vectors and matrices is fixed, we must reduce the maximum number of controlled modes to do so.
This is an interesting possibility: getting a temporal control with additional degrees of freedom with respect to a leaky integrator on a limited number of modes (ideally the ones affected by larger turbulence and vibration disturbances), allowing, in theory, for a significantly better correction. 

In the following paragraph we present how a mixed-control approach (i.e., IIR filters for a few modes and integrator for all the other modes) can be implemented.
This is similar to what has been presented in \cite{agapito:11} for an observer-based control.

We start from Eq. \ref{eq:stateIIR} and \ref{eq:stateINT}, and considering that $\boldsymbol{B}=\mathcal{G}(g) \, \mathcal{B}$ where $\mathcal{B} \in \mathbb{R}^{N_{\mathrm{f}} \times 1}$ is a matrix 
without dependencies on the filter gain\footnote{A filter can  always be expressed as product between its gain and an unitary gain filter.} and $\mathcal{G}(g) \in \mathbb{R}^{N_{\mathrm{f}} \times N_{\mathrm{f}}}$ is a diagonal matrix function of the filter gain $g$ \footnote{This decomposition of the input matrix $\boldsymbol{B}$ is done to get an expression compatible with the $\boldsymbol{\Delta}$ and $\boldsymbol{\Xi}$ matrices so the capability of updating the filter gains during closed loop operations.}, RTC matrices become:
\begin{equation} \label{eq:RTCm1}
\boldsymbol{\Lambda} := \left[ \begin{array}{cc}
\boldsymbol{A} & 0\\
0 & \boldsymbol{I}_{N_{\mathrm{I}}}\\
\end{array} \right]
\end{equation}
\begin{equation} \label{eq:RTCm2}
\boldsymbol{\Xi} := \left[ \begin{array}{ccc} \mathcal{B} & 0 & 0\\ 0 & \boldsymbol{I}_{N_{\mathrm{I}}} & 0\\
\end{array} \right] \mathcal{R}
\end{equation}
\begin{equation} \label{eq:RTCm3}
\boldsymbol{\Delta} := \left[ \begin{array}{cc}
\mathcal{G}(g) & 0\\
0 & \boldsymbol{G}\\
\end{array} \right]
\end{equation}
\begin{equation} \label{eq:RTCm4}
\boldsymbol{\Sigma} := \mathcal{M}\left[ \begin{array}{cc}
\boldsymbol{C} & 0\\
0 & \boldsymbol{I}_{N_{\mathrm{I}}} \\
0 & 0
\end{array} \right]
\end{equation}
So, it is clear that larger state dimension $N_{\mathrm{f}}$ means less modes that can be controlled.
Note that this can be easily extended to any number of modes controller with IIR filters, provided that ${N_{\mathrm{I}}} + \sum_i{(N_{\mathrm{f}})_i} \leq 672$.

Following this mixed-control approach, Tip/Tilt filters have been optimized to reject LBT vibrations, using $N_{\mathrm{f}}=8$ (that means $N_{\mathrm{\beta}}=N_{\mathrm{\alpha}}=4$).
However, for other modes a leaky integrator was selected ($N_{\mathrm{f}}=1$) and only its 1-element state update matrix, that is its pole, was optimized \cite{Agapito2019}.


Note that leaky integrators are a fundamental aspect of our approach for the SOUL control system, but we will not focus on them here: we direct the reader to  \cite{Agapito2019} where an analysis of the advantage of this kind of control with respect to the simple integrator is reported along with a description of the forgetting factors optimization procedure.
As shown in that article, leaky integrators are able to reduce noise and aliasing propagation, in particular for high order modes, improving performance, image contrast and robustness.

\subsection{Optimization of Tip/Tilt Control Filters}\label{sec:TTcontrol}

\begin{figure}
    \begin{center}
    \begin{tabular}{c}
        \includegraphics[width=0.45\textwidth]{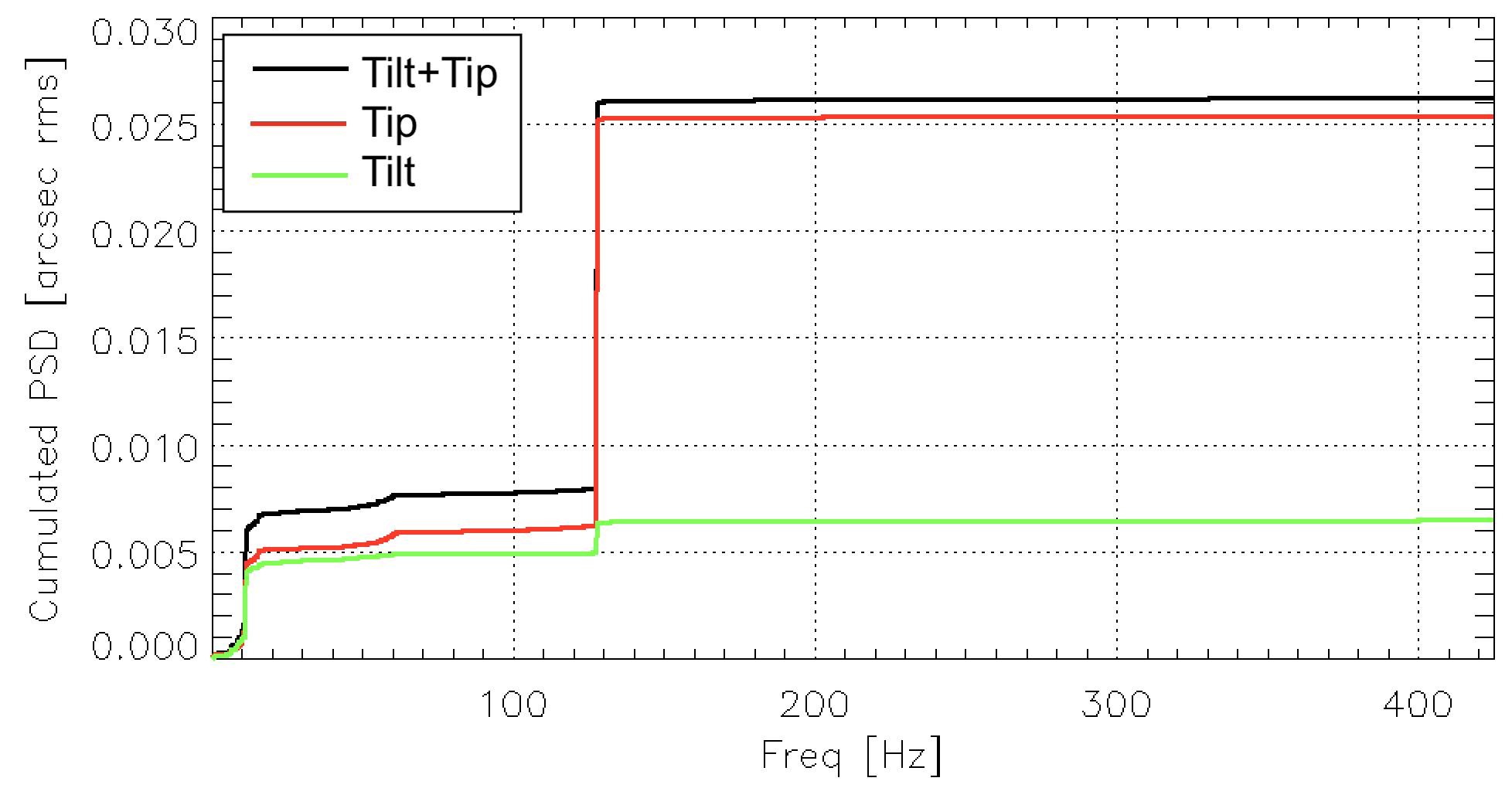}
    \end{tabular}
    \end{center}
	 \caption{\label{fig:vibration} Cumulated Tip/Tilt jitter from residual modes showing strong lines at 11 and 127Hz.}
\end{figure}

\begin{figure}
    \begin{center}
    \begin{tabular}{c}
        \includegraphics[width=0.425\textwidth]{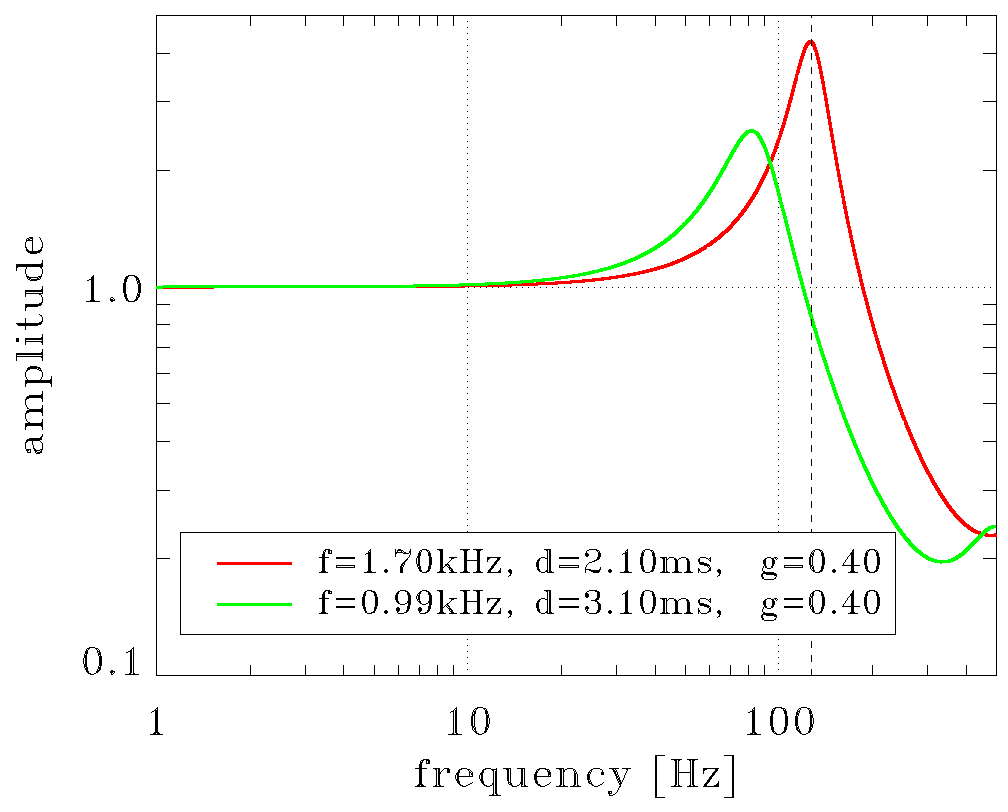}
    \end{tabular}
    \end{center}
	 \caption{\label{fig:TFcomp} Complementary sensitivity function (known also as noise transfer function in the AO field) of FLAO and SOUL system  with an integrator control with gain 0.4 at maximum framerate.}
\end{figure}

Tip and Tilt are distinctive because of strong telescope structure vibrations: on the LBT they have a principal component around 13Hz and, like other telescopes, these vibrations affect mainly Tip/Tilt \cite{10.1117/12.925984}.
If not corrected properly they are one big limitation for the performance.
Unfortunately, during the commissioning of SOUL, we noted a new Tip/Tilt vibration: when integrator gains are above a certain level, a strong vibration at 127Hz appears (see Fig.\ref{fig:vibration}).
We found a control-structure interaction of the ASM:
the dynamics of the mirror actuators interacting with the mirror-supporting structure induced a vibration.
Previously this had never appeared because the control of FLAO has its bandwidth below 127 Hz, since it has a significant longer minimum delay on Tip/Tilt: 3.1ms with respect to 2.1ms for SOUL.
Thanks to this lower delay the complementary sensitivity function of SOUL has its bandwidth at -3dB well above 127Hz and for a gain of 0.4 the amplitude at this frequency is greater than 4 (see Fig. \ref{fig:TFcomp}).
Hence, SOUL is able to excite a previously unknown resonance of the ASM.
This greatly limits the range of integrator gains that can be used, and, consequently, the overall performance of the system: in practice, with an integrator control, we are forced to reduce bandwidth on Tip/Tilt, going back to values close to the ones of FLAO.
For example, at a framerate of 1700Hz we are forced to maximum gains of about 0.2, getting the same bandwidth of FLAO with a framerate of 1000Hz and a gain of 0.4 (a gain close to the stability limit but still usable).
Fortunately this issue can be solved at the level of the AO control system without acting on the ASM hardware.

Control filters for Tip/Tilt have been optimized using semi-analytical routines described in \cite{10.1117/12.925896} and \cite{10.1117/1.JATIS.5.4.049001}.
A cost function considering temporal error and disturbance propagation has been used.
As additional inputs we used a typical LBT vibration PSD \cite{10.1117/12.925984} boosted by a 2mas RMS at 127Hz.
The RMS of the 127Hz peak is simply used to avoid a high transfer function amplitude at this frequency and it is not directly related to the actual RMS of such a vibration.
During the optimization we impose a pole in z=1 to guarantee an integral action so the ability to track a constant set-point and we impose a stability constraint as described in \cite{10.1117/12.925896}. But instead of a condition on the roots of the characteristic polynomial, in this case we verify that:
\begin{equation}
    \|H(z)P(z)-1\|>\epsilon
\end{equation}
where $H$ is the transfer function of the control filter, $P$ is the transfer function of the plant and $0<\epsilon<1$ is an arbitrary scalar value. So, the minimum distance from -1 in the Nyquist plot must be greater than $\epsilon$.
This satisfies at the same time two conditions of the stability margins: gain margin $>1/(1-\epsilon)$ and phase margin $>2\sin^{-1}(\epsilon/2)$, but it is more stringent than the union of the two.
In our computation we choose $\epsilon=0.2$.

\begin{figure}
    \begin{center}
    \begin{tabular}{c}
        \includegraphics[width=0.45\textwidth]{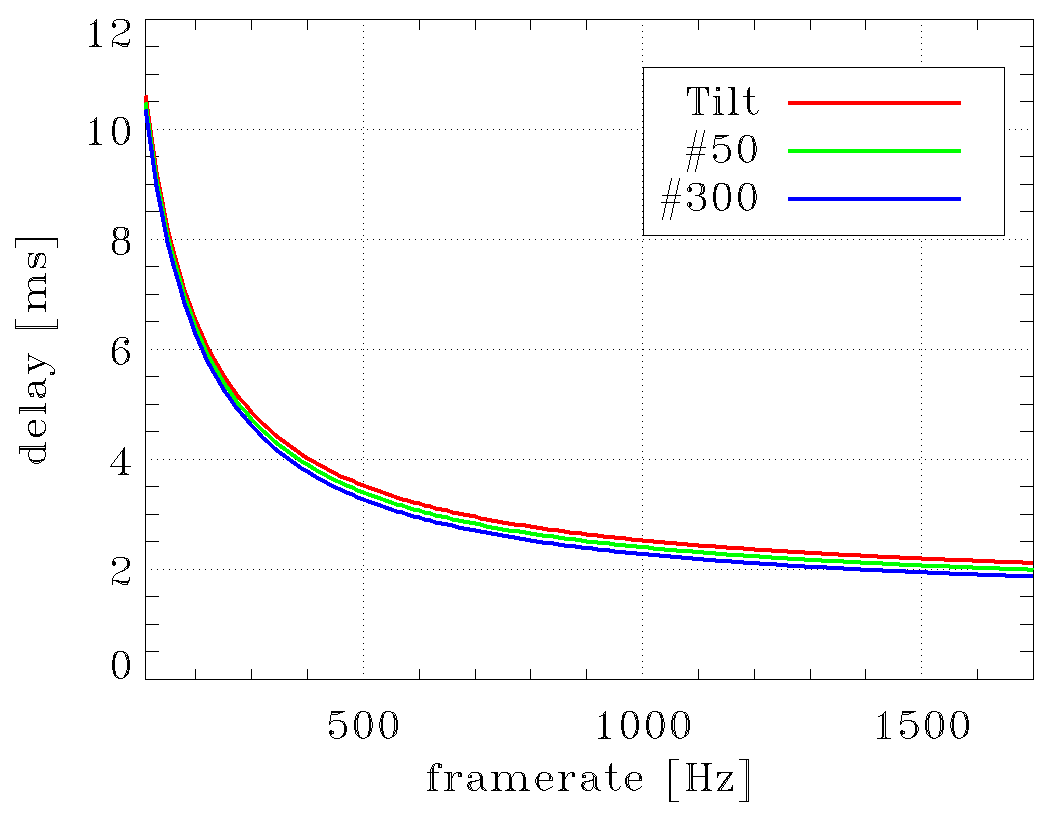}
    \end{tabular}
    \end{center}
	 \caption{
	 \label{fig:delay} SOUL system total delay as a function of framerate. The values can slightly differ from mode to mode. This difference come from the ASM behavior.}
\end{figure}

Note that the plant $P$ is modelled as a pure delay, and this delay has been computed analytically \cite{10.1117/12.2234444} and verified in controlled conditions measuring the closed loop transfer function resonance peak frequency.
Its value is shown in Fig. \ref{fig:delay} and slightly differs from mode to mode due to the ASM behavior as described in \cite{Riccardi2008}.
%

\begin{figure}
    \begin{center}
        \subfigure[Sensitivity function.
        \label{fig:IIR1700RTF}]
        {\includegraphics[width=0.9\columnwidth]{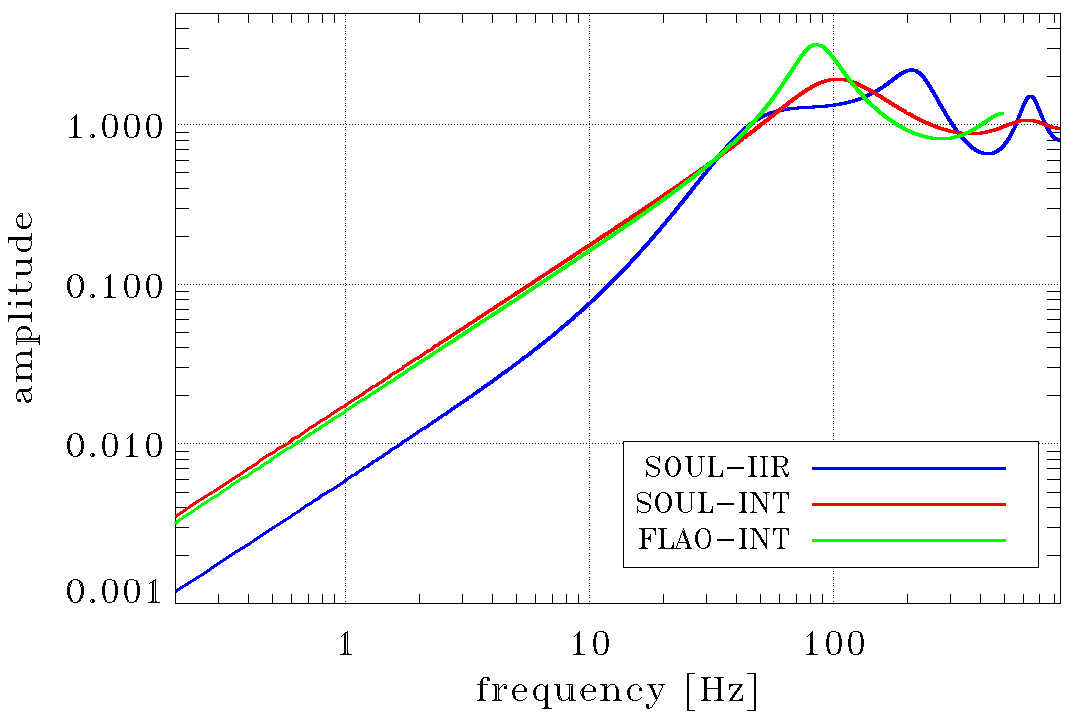}}
        
        \subfigure[Complementary sensitivity function.
        \label{fig:IIR1700NTF}]
        {\includegraphics[width=0.9\columnwidth]{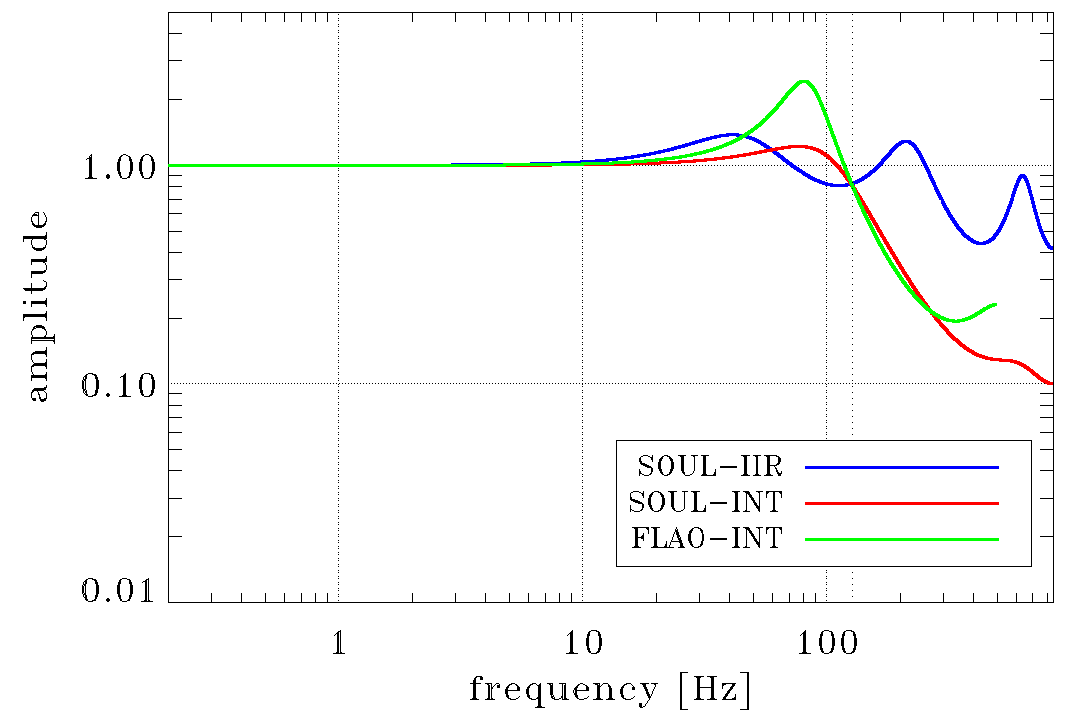}}
    \end{center}
	 \caption{\label{fig:IIR1700TF} Transfer function amplitude. 
	 SOUL with IIR filter control is compared with SOUL and FLAO with integrator (INT in the legend) control. 
	 SOUL framerate is 1700Hz, FLAO one is 1000Hz and gains are 0.69, 0.21, and 0.39 for SOUL-IIR, SOUL-INT and FLAO-INT respectively.
	 }
\end{figure}

We optimized several filters for different frame rates, from 750 to 1700Hz (see Tab. \ref{Tab:roots}).
We stopped at 750Hz because we saw, in numerical simulations, that at lower frame rates the $N_{\mathrm{f}}=8$ filter we were optimizing gave a performance similar to a pure integrator.
The amplitude of the closed loop Transfer Functions (TF) with the IIR filter for 1700Hz framerate are shown in Fig. \ref{fig:IIR1700TF}.
Here they are compared to closed loop TFs with an integrator and with ones with FLAO parameters (framerate 1000Hz and delay 3.1ms).
We selected gains that give the same complementary sensitivity function amplitude at 127Hz and, in particular for SOUL with an integrator, a gain of 0.21 that proves not to excite the resonance of the ASM. 
Note that the actual gain values could be different 
because the filter gain will be optimized on the fly.
The resonance peak of the TF with the IIR filter is moved to a higher frequency, so that it does not excite the 127Hz resonance of the ASM and the amplitude below 30Hz is lower, so that it can better reject atmospheric turbulence and the ASM spider resonance at 13Hz.
Moreover, the IIR filter has good robustness: phase and gain margins are 49$^{\circ}$ and 1.8 respectively. 
\begin{table}
\caption{Zeros, $\boldsymbol{\beta}$, and poles, $\boldsymbol{\alpha}$, of the IIR filters (see Eq. \ref{eq:tfIIR}) as a function of the frame rate (that is the inverse of the integration time of the detector)}.
\label{Tab:roots}
\begin{center}
\begin{small}
	\begin{tabular}{|l|c|c|c|c|}
		\hline
		frame rate [Hz] & \multicolumn{4}{c}{zeros} \\
		\hline
		1700 & -0.310 & 0.690 & 0.698 & 0.705 \\
		1350 &  0.322 & 0.596 & 0.601 & 0.608 \\
		1000 &  0.337 & 0.345 & 0.345 & 0.353 \\
		 750 &  0.316 & 0.316 & 0.329 & 0.329 \\
		\hline
		frame rate [Hz] & \multicolumn{4}{c}{poles} \\
		\hline
		1700 &  0.042 &  0.102 & 0.954 & 1.000 \\
		1350 &  0.285 &  0.289 & 0.912 & 1.000 \\
		1000 & -0.187 & -0.187 & 0.852 & 1.000 \\
		 750 & -0.461 &  0.633 & 0.720 & 1.000 \\
		\hline
	\end{tabular}
\end{small}
\end{center}
\end{table}

\subsection{Modal Gain Machine}\label{sec:MGM}

As stated before, modal gains can be updated on the fly, but an optimization algorithm is required to benefit from this feature.
In 2010, when FLAO went on sky, a trial-and-error procedure was implemented.
This procedure spans a large range of gains in a short time interval (a few seconds) before the science observation starts.
The merit function used is the slope RMS and three values are optimized: gains of Tip/Tilt, modes between 2 and 120 and modes from 120 to the last one.
Then, these gains were used for the whole observation.
This rough procedure is quite effective for a WFS like the pyramid where the sensitivity changes as a function of the AO residual: in this case a trial-and-error strategy avoids the requirement of estimating the current sensitivity, the so-called optical gain  \cite{2008SPIE.7015E..54K,2008ApOpt..47...79K,EspositoNCPA2015,Esposito2020}.
It also has its drawbacks, in particular it is limited to a few sets of modes and no on-line update of the gain is expected.
Then, in 2015 \cite{EspositoNCPA2015,Esposito2020} an optical gain estimation algorithm named ``Optical Gain Tracking Loop'' (OGTL) was added to the FLAO control so the unitary gain of the PWFS was restored.
Hence, we started to develop a modal control optimization-like \cite{GENDRON1994} algorithm for the PWFS. 

This algorithm is called Modal Gain Machine (MGM).
Its purpose is to optimize the gains of the filters (not only integrators, but any kind of filters) every few seconds to keep them optimized as a function of the current disturbance status (not only atmosphere, but also vibrations).
The minimization criterion and cost function are derived from  \cite{1998ApOpt..37.4623D} with a few modifications to be compatible with the PWFS and they respectively are:
\begin{equation}\label{eq:minCriterion}
    \hat{g}_i = \min_{g_i}{J_i}
\end{equation}
and:
\begin{equation}\label{eq:cost}
    J_i = \sum^{1/2T}_{f=1/nT}{\Phi_i^{\mathrm{meas}}(f)} = \sum^{1/2T}_{f=1/nT}{\| W_i(z,g_i)\|^2 \Phi_i^{\mathrm{pol}}(f)}
\end{equation}
where $i$ is the mode index, $\Phi^{\mathrm{meas}}(f)$ is the PSD of the  measurement, $\Phi^{\mathrm{pol}}(f)$ is the PSD of the pseudo-open loop and $W(z,g_i) = (1+H(z,g)P(z))^{-1}$.
Note that $H(z,g) = \boldsymbol{C}(\boldsymbol{I}-\boldsymbol{A}z^{-1})^{-1} \mathcal{G}(g) \, \mathcal{B}$ is a linear function of the gain $g$ so a simple minimization algorithm can be used to find the optimal gain in the range of stable gains. 
This range is typically reduced by 10\% to avoid cases close to a null phase margin.

In the case of a non-linear WFS like the PWFS, the plant can be approximated as
\begin{equation}\label{eq:plant}
    P(z)=\gamma_i z^{-d} \; ,
\end{equation}
where $\gamma_i$ is a scalar value that models the loss of wavefront sensor sensitivity, also known as optical gain, for the $i$-th mode and $d$ is the total delay.
Then, pseudo-open loop modes $\boldsymbol{m}_{\mathrm{pol}} \in \mathbb{R}^{N}$, that are used to compute $\Phi^{\mathrm{pol}}(f)$, can be defined as:
\begin{equation}\label{eq:pol}
    \boldsymbol{m}_{\mathrm{pol}}(z) = \boldsymbol{m}(z) + \boldsymbol{\Gamma}^{-1} \mathcal{R} \boldsymbol{s}(z) z^{-d} \; ,
\end{equation}
where $\boldsymbol{m} \in \mathbb{R}^{N}$ is the modal command vector, $\mathcal{R}$ is the reconstruction matrix and $\boldsymbol{\Gamma} \in \mathbb{R}^{N \times N}$ is a diagonal matrix whose diagonal elements are the $\gamma_i$ values previously cited.

Note that considering the optical gain in the model of the plant is important because otherwise optimization might greatly underperform and stability might be compromised due to the model error. In fact, PWFS sensitivity in partial correction regime (\cite{2001A&A...369L...9E}) is always different from the nominal one (see \cite{2018SPIE10703E..20D,10.1117/1.JATIS.5.4.049001}).

OGTL at LBT is not simply estimating the optical gain, but it is also scaling the slope vector $\boldsymbol{s}$ by a sensitivity scalar value $k$ to enables the application of the proper Non-Common-Path Aberration (NCPA) offset values as it is described in \cite{Esposito2020}.
We are interested in this feature because it changes the effective loop gain, that becomes the product of WFS optical gain $\gamma_i$, the sensitivity factor $k$, and filter gain $g_i$.
So we updated Eq. \ref{eq:plant} and Eq. \ref{eq:pol} replacing $\gamma_i$ with $\chi_i=\gamma_i k$ and  $\boldsymbol{\Gamma}$ with $\boldsymbol{X}=\boldsymbol{\Gamma} k$.
Note that 
OGTL restores the unitary gain, $\chi_i = 1$, for the mode where the probe signal is applied (typically mode no. 30).
The effective loop gain for such mode is $g_i$ and Eq. \ref{eq:plant} is a pure delay, $z^{-d}$.
Considering this we tabulated $\chi_i$ as a function of the correction level while the optical gain coefficients (used to derive $\chi_i$ values) are computed as shown in \cite{10.1117/1.JATIS.5.4.049001}.

Finally, the gain value applied in the RTC, $g_{i}^{RTC}$, is computed as:
\begin{equation}\label{eq:gainUpdate}
    g_{i}^{\mathrm{RTC}}(\kappa) = (1-\rho)g_{i}^{RTC}(\kappa-1) + \rho \nu \hat{g}_{i}(\kappa)
\end{equation}
where $\kappa$ is the optimization step, and $\nu<1$ and $\rho<1$.
Both parameters improve robustness: $\nu$ reduces the applied absolute value, while the low pass filter ${\rho}/{(1-(1-\rho)z^{-1})}$ reduces noise propagation and quick dynamic variations.
In fact, we are interested in a control robust against modeling errors, for example on $\chi_i$ and $d$ values, but also on the calibration errors of the reconstruction matrix and, consequently, on the pseudo-open loop modes.
Typical values used are $\nu=0.9$ and $\rho=0.5$.

MGM coordinates with OGTL to avoid unstable behaviour because changes of $g$ and $k$ affect the effective loop gain.
Hence, the steps of the algorithm are:
\begin{enumerate}
    \item record telemetry data
    \item estimate optical gain (as shown in  \cite{Esposito2020})
    \item estimate correction level then, for each mode $i$, retrieve $\chi_i$
    \item compute the pseudo open loop modes time history $\boldsymbol{m}_{\mathrm{pol}}$
    \item compute the PSDs of the pseudo open loop modes $\Phi^{\mathrm{pol}}(f)$
    \item for each mode $i$ compute the cost function $J_i$ and find the gain $\hat{g}_i$ that minimizes it 
    \item set the new sensitivity value $k$ and modal gain vector $\boldsymbol{g}^{\mathrm{RTC}}$ scaled by $k({\kappa-1})/k({\kappa})$.
\end{enumerate}

This last step is required to allow both MGM and OGTL to work with a same data set with a fixed effective loop gain (so with fixed $k$ and $g$).
Then, scaling by $k({\kappa-1})/k({\kappa})$ allows to correctly apply the effective loop gain determined by MGM, $\boldsymbol{g}^{\mathrm{RTC}}(\kappa)\boldsymbol{\Gamma} k(\kappa-1)$, instead of $\boldsymbol{g}^{\mathrm{RTC}}(\kappa)\boldsymbol{\Gamma} k(\kappa)$.
Here, we have considered that the correction level and, consequently, $\Gamma$ changes in a much longer time than one optimization step \footnote{Correction level depends on seeing, that typically evolves slower than the few seconds required by an optimization step (5-10s), and on effective loop gain that slowly changes thanks to the low pass filter in Eq. \ref{eq:gainUpdate}.}.

The correction level can be estimated from telemetry data as shown in  \cite{2004SPIE.5490..118F}, but currently this part of the algorithm has not been implemented. So we use a set of $\chi_i$ values from a specific condition: 0.8arcsec seeing.
This is not ideal, but it proved to guarantee stability together with the other safety margins, and, as we wrote before, we are more interested in robustness than absolute performance.

The algorithm is implemented in Python and it is triggered periodically on the PWFS workstation: the period is a configurable value that we set to one second as a good compromise between accuracy and computational load. 
The SOUL software running on the PWFS workstation is based on a set of processes communicating with each other and with the ASM software using shared memory or message passing mechanism, depending on the specific case. 
For more details about the SOUL software infrastructure refer to  \cite{Rossi2019}.
In the implementation of MGM we made an effort to extend the SOUL software in a modular way.
As a starting point we designed a base class called \emph{auxloop} to model an additional process that can run in parallel with the rest of the SOUL software and interface with SOUL RTC by reading from the live telemetry data stream and writing to the RTC parameters.
Subclasses of an \emph{auxloop} can be used to perform specific tasks requiring the continuous configuration of the RTC parameters or to compute relevant information derived from telemetry which can then be displayed live to the operators during the observation and/or stored in our monitoring database.
In fact, besides the one for MGM, we have developed other \emph{auxloop} specializations performing tasks such as the computation of the vector of the RMS of the modal residuals or the current Strehl ratio from the WFS slopes or the estimated magnitude of the current guiding star.
The subclass that implements MGM is called \emph{optModalGainTask}. 
In order to improve the performance of the most demanding parts of our algorithm, like the minimization of the cost function to find the optimal gain value, we relied on the \emph{multiprocessing} Python package to parallelize the computation on a mode-by-mode basis.

\section{Results}\label{sec:results}

\begin{table}
\caption{Summary of IIR filter and integrator (INT) performance and parameters. SR is the average value computed on sets of 3 (daytime) and 19 (on-sky) images with 9.4s (daytime) and 6.2 (on-sky) exposure time. Jitter is the average RMS value computed on sets of 3 (daytime) and 19 (on-sky) telemetry data acquisitions of 4000frames. Note that daytime images are affected by not-compensated LUCI1 NCPA of about 100nm that reduce significantly their SR.}
\label{Tab:TT}
\begin{center}
\begin{small}
	\begin{tabular}{|l|l|c|c|c|c|c|}
		\hline
		 & \multirow{2}{*}{contr.} & H band & jitter & freq.  & \multirow{2}{*}{gain} & reject.\\
		 &  & SR & [mas] & [Hz] &  & @13Hz\\
		\hline
		\multirow{2}{*}{Daytime} &
		IIR & 0.67 & 7 & 1000 & 1.03 & 0.10 \\
		 & INT & 0.51 & 11 & 1000 & 0.36 & 0.23 \\
		\hline
		\multirow{2}{*}{On-sky} &
		IIR & 0.72 & 8 & 1700 & 0.69 & 0.11 \\
		 & INT & 0.49 & 12 & 1700 & 0.20 & 0.24 \\
		\hline
	\end{tabular}
\end{small}
\end{center}
\end{table}
In this section, we show a sample of results obtained both in daytime, with the calibration source, and on sky, during the commissioning nights. All reported tests have been performed on the SOUL system coupled with the spectro-imager LUCI1 \cite{2010SPIE.7735E..7WS,2012SPIE.8446E..5LB,2018SPIE10702E..0BH}. We used LUCI's high resolution imaging camera (N30), providing a plate scale of 0.015asec/pix.    We first compare IIR filters with integrators and then MGM with "old" gain optimization.

\subsection{IIR filter results}

In the tests reported here below, the loop gains are optimized with the trial-and-error method, OGTL is active, and MGM is not used.

During 2019, we tested the Tip/Tilt IIR filters, using a calibration source (daytime tests) and a retro-reflector positioned at the short focus of the ASM \cite{2010ApOpt..49G.174E}. 
In this case no turbulence was simulated by the ASM, and only telescope structure vibrations were present.
We closed the loop at 1000Hz and then we optimized the loop gains. At first, we used IIR filters, then the integrators and we acquired PSFs at 1650nm with the LUCI imager.
As can be seen in Tab. \ref{Tab:TT}, IIR filters give a steady performance of 67.3$\pm$1.7\% H band SR \footnote{average $\pm$ standard deviation values computed on a set of 3 images with 9.4s exposure time.}, while integrators give a performance of 51.0$\pm$6.8\% H band SR under the same conditions.
We underline that all SR values are depressed due to the LUCI1 NCPA of about 100nm that, in this case, are not compensated.
This is the reason why these SRs are below 70\%. 
However, the IIR filter always outperforms the integrator.
Analyzing the AO telemetry data, we find the residual Tip/Tilt jitter for IIR filters and integrators to be 7 mas and 11 mas respectively.
In both cases the 127Hz vibration is not excited, but, while the gain optimization finds the optimal value at 1.03 for the IIR \footnote{A gain > 1 is generally known to be unstable for the classical integrator control, however the loop with this kind of filter is stable with gains up to 2.}, in the case of the integrator the optimal gain is limited to 0.36.
The ratio between these gains, about a factor 3, is in line with what is shown in Sec. \ref{sec:TTcontrol} and, as expected, allows the IIR filter for a better rejection of the 13Hz vibration on Tip/Tilt (0.10 with respect to 0.23 of the integrator) while giving the same amplitude of the complementary sensitivity functions at 127Hz (between 0.8 and 0.9).
Both controls with such gains have good robustness properties: phase and gain margins are 40$^{\circ}$ and 2.14 for the IIR filter and 51$^{\circ}$ and 2.42 for the integrator.

\begin{figure*}
    \begin{center}
    \begin{tabular}{c}
        \includegraphics[width=0.9\textwidth]{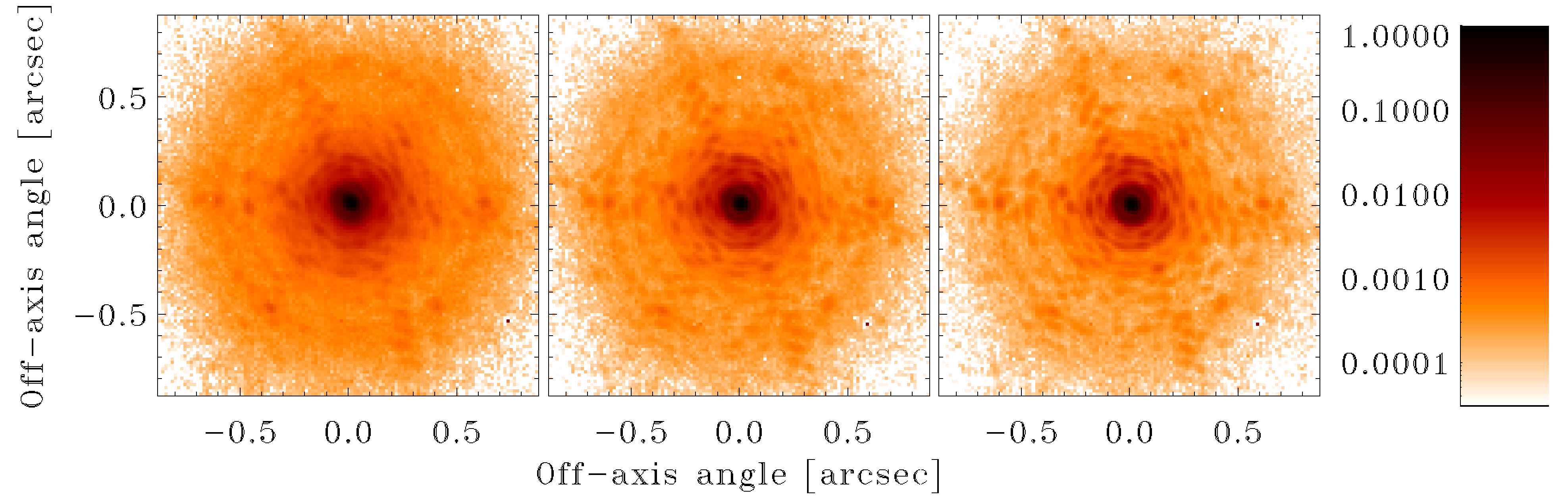}
    \end{tabular}
    \end{center}
	 \caption{\label{fig:PSFnightIIR} A sample of 3 images (sub-frame cut) of a sequence of 19 of a bright NGS acquired at 1646nm on LUCI1 (6.2s exposure time) with IIR filters on Tip/Tilt. The NGS is a R 9.7 star, the loop frame rate is 1700Hz and 40$\times$40 sub-apertures are used. An average of 20ph/sa/frame (ph is photon and sa is sub-aperture) are detected on the WFS valid sub-apertures (note that a sub-aperture is the set of four pixels, one per pupil image). Peaks of the images are normalized to 1. From left to right, we report the one with lowest SR (0.58), the one with SR close to average (0.72) and the one with highest SR (0.79). During the acquisition of the sequence, the DIMM seeing varies from 1.06 to 1.58arcsec on the line of sight and average wind speed estimated from telemetry data from 6 to 10m/s. None of the 3 images show signature of vibrations (speckle elongation), while the PSF contrast clearly increases together with the SR as expected for improving seeing conditions.}
\end{figure*}

\begin{figure*}
    \begin{center}
    \begin{tabular}{c}
        \includegraphics[width=0.89\textwidth]{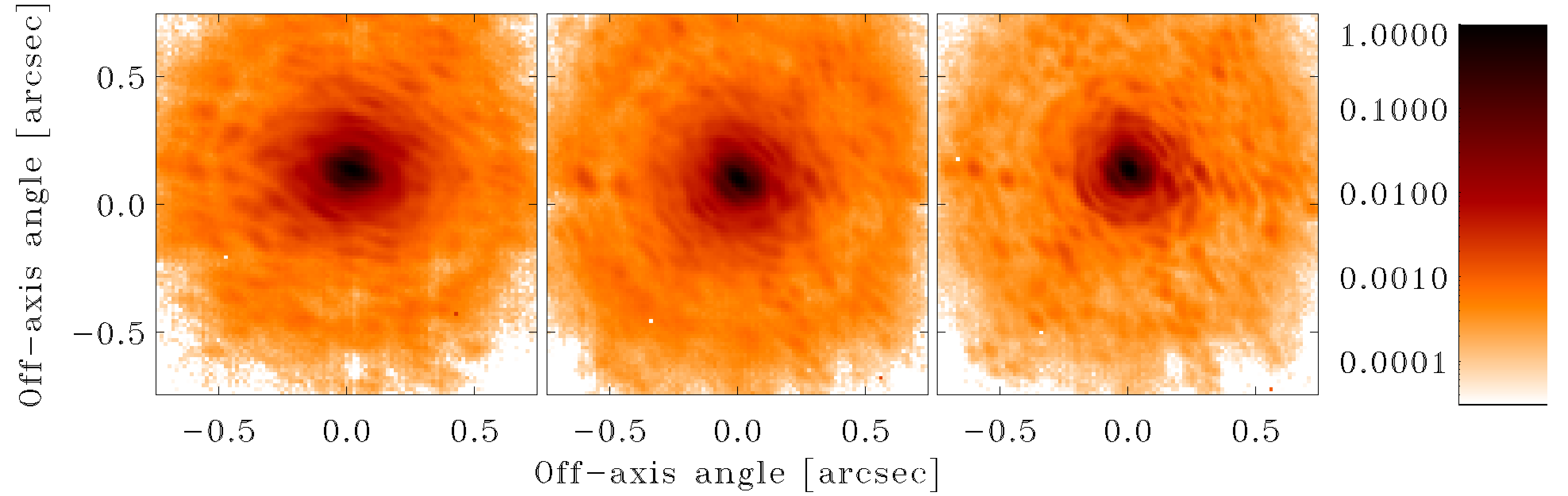}
    \end{tabular}
    \end{center}
	 \caption{\label{fig:PSFnightINT} As in fig.\ref{fig:PSFnightIIR}, a sample of 3 images of 19 acquired on the NGS (9.4s exposure time) with integrators on Tip/Tilt. The NGS is a R 9.7 star, the loop frame rate is 1700Hz and 40$\times$40 sub-apertures are used. An average of 20ph/sa/frame are detected on the WFS valid sub-apertures. Peaks of the images are normalized to 1. From left to right, we report the one with lowest SR (0.36), the one with SR close to average (0.49) and the one with highest SR (0.60). The DIMM seeing varies from 1.07 to 1.56arcsec during the acquisition and average wind speed estimated from telemetry data from 8 to 12m/s. All the 3 images show clear signature of residual vibrations, more pronounced for the case at lower SR. The direction of the vibration rotates in the frame due to the field de-rotation of the instrument.}
\end{figure*}

During night time, we compared integrators and IIR filters on bright guide stars (R<10) with a frame rate of 1700Hz.
These tests, as further commissioning data did, confirmed the ability of the IIR filter to provide a good rejection of telescope vibrations, avoiding the excitation of the 127Hz resonance.
Here, as an example, we select two sets of 19 PSFs acquired on the LUCI1 at 1650nm (FeII filter) with seeing varying from 1.0 to 1.6arcsec.
As can be seen in Tab. \ref{Tab:TT}, the SRs measured on long exposures with LUCI1 are in the  57.9 - 79.5\% range (see Fig. \ref{fig:PSFnightIIR}) and Tip/Tilt RMS jitter is 7 - 11mas with IIR filters, while with integrator they span respectively from 36.0 to 59.6\% (see Fig. \ref{fig:PSFnightINT}) and from 10 to 16mas.
The expected performance from simulations (see \cite{Pinna2019}) with such seeing values is between 50 and 80\% and is in good agreement with the values obtained with IIR filters.
Hence, the nominal performance is retrieved.

\begin{figure}
    \begin{center}
        \subfigure[Pseudo-open loop. Note that jitter is not the same in the two cases because the vibration level is varying in time due to wind load and instruments pumps/fans status. Here, in particular, we see a change in the level of 13Hz vibration that is due to the spider arm resonance and it is sensitive to wind load.
        \label{fig:IIRjitterNight_open}]
        {\includegraphics[width=0.9\columnwidth]{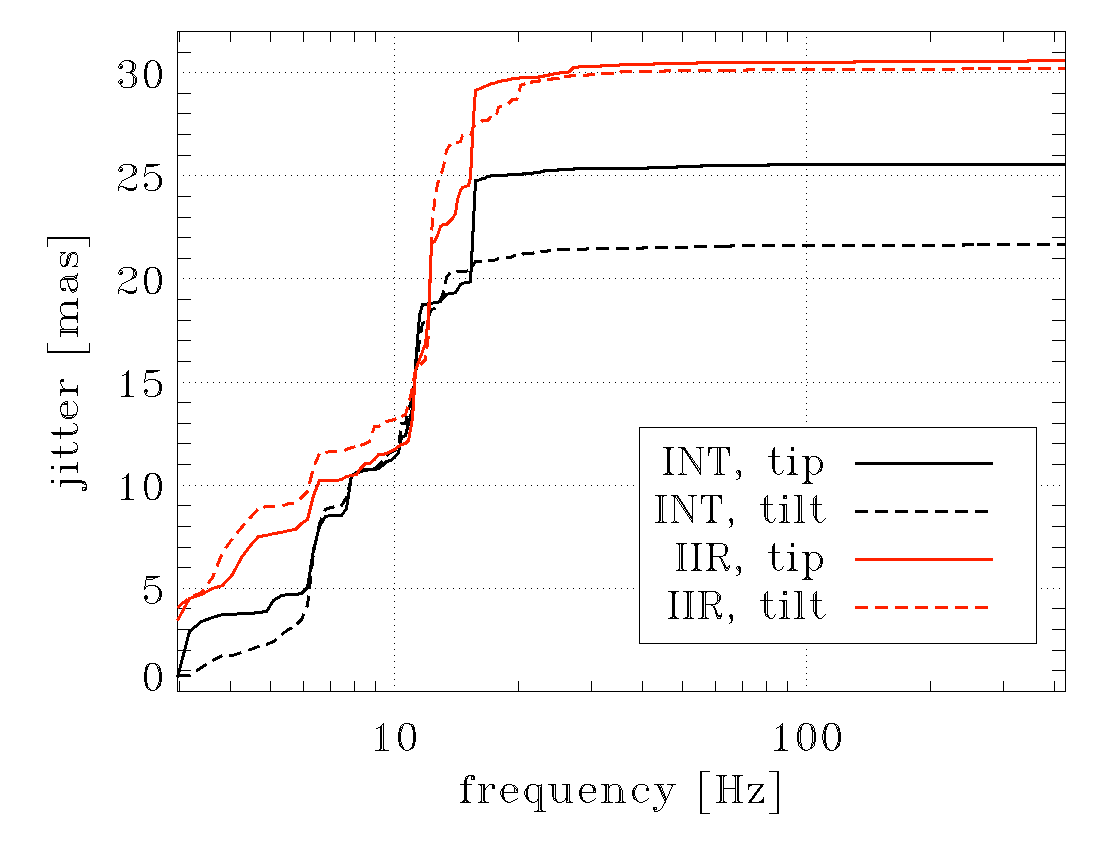}}
        
        \subfigure[Residual.
        \label{fig:IIRjitterNight_close}]
        {\includegraphics[width=0.9\columnwidth]{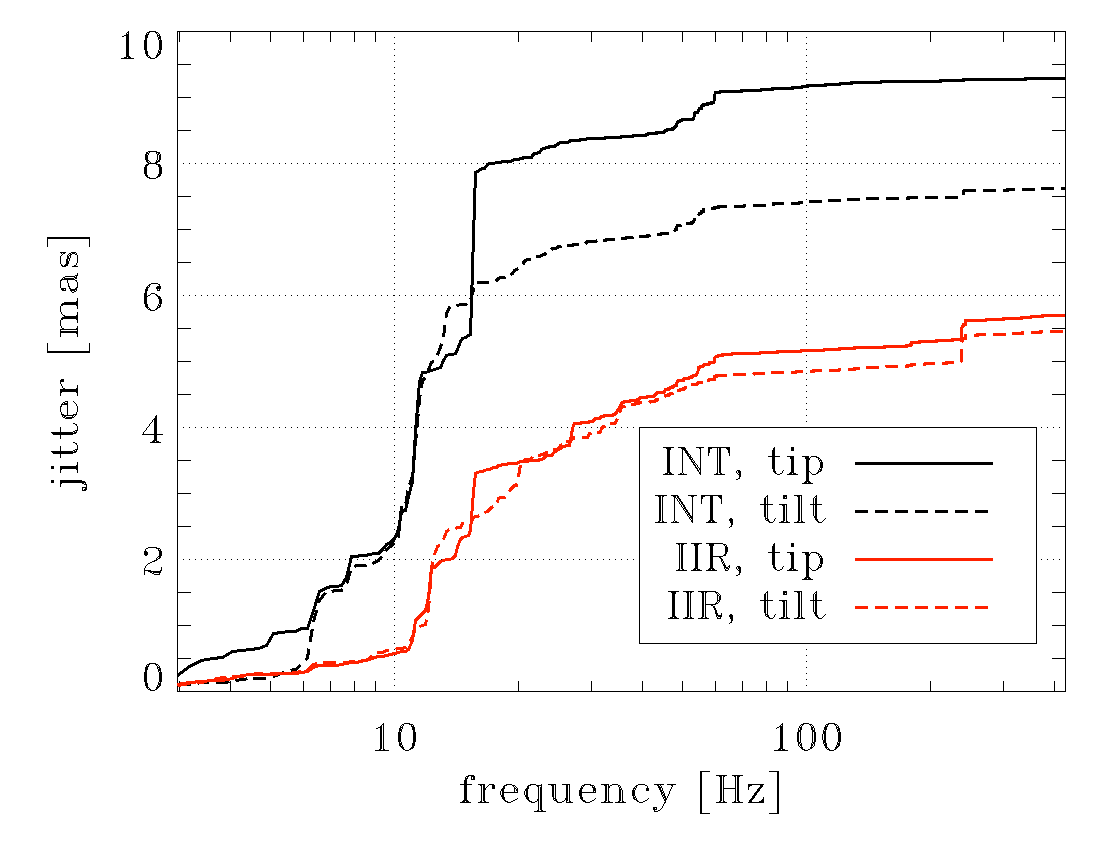}}
    \end{center}
	 \caption{\label{fig:IIRjitterNight} On-sky Tip/Tilt cumulated jitter from 5s long data sets, comparison between IIR filters and integrators starting from 3Hz.
	 We select data sets with residual jitter close to the average values: 7.9mas and 12.3mas for IIR filters and integrators respectively (instead pseudo-open loop values in the range 3-425Hz are 43 and 34mas).}
\end{figure}

Note that the optimized Tip/Tilt gains are 0.20 for Integrator and 0.69 for IIR filter: they are close to the values used to plot TFs in Fig. \ref{fig:IIR1700TF} and the same considerations presented in Sec. \ref{sec:TTcontrol} are valid.
As for the test with the calibration source, both controls are robust: phase and gain margins are 49$^{\circ}$ and 1.82 for the IIR filter and 51$^{\circ}$ and 2.34 for the integrator.
In Fig. \ref{fig:IIRjitterNight}, we compare both pseudo-open loop and residual cumulative spectra for the Tip/Tilt of IIR and integrator cases between 3 and 425Hz.
Here we can see that, even in presence of higher input (vibration level is varying in time due to wind load and instruments pumps/fans status), the residuals produced by IIR filter are significantly lower than those provided by the integrator.
As described in Sec. \ref{sec:TTcontrol} this demonstrates the better rejection achieved for frequencies <30Hz, without exciting the 127Hz resonance.
\subsection{MGM results}

\begin{figure}
    \begin{center}
        \subfigure[Daytime. Framerate is 990Hz. Black line, gains optimized with MGM, red line, gains optimized by trial-and-error method.
        \label{fig:GAINday}]
        {\includegraphics[width=0.9\columnwidth]{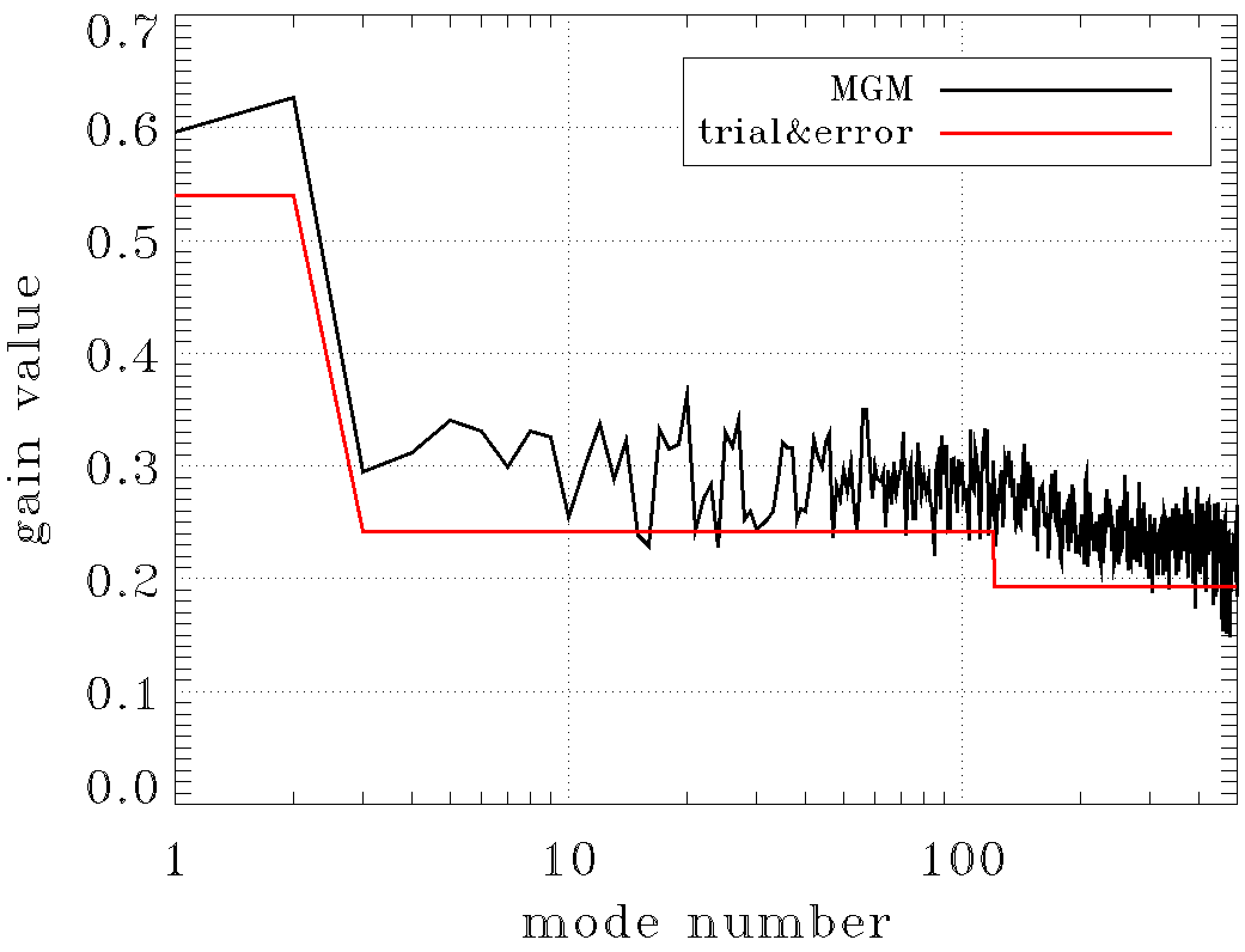}}
        
        \subfigure[On-sky optimized gain vectors at a time distance of 1 minute. Frame rate is 1212Hz.
        Note: trial-and-error method optimized gains of 0.387, 0.198 and 0.164 for the three sets of modes respectively.
        \label{fig:GAINnight}]
        {\includegraphics[width=0.9\columnwidth]{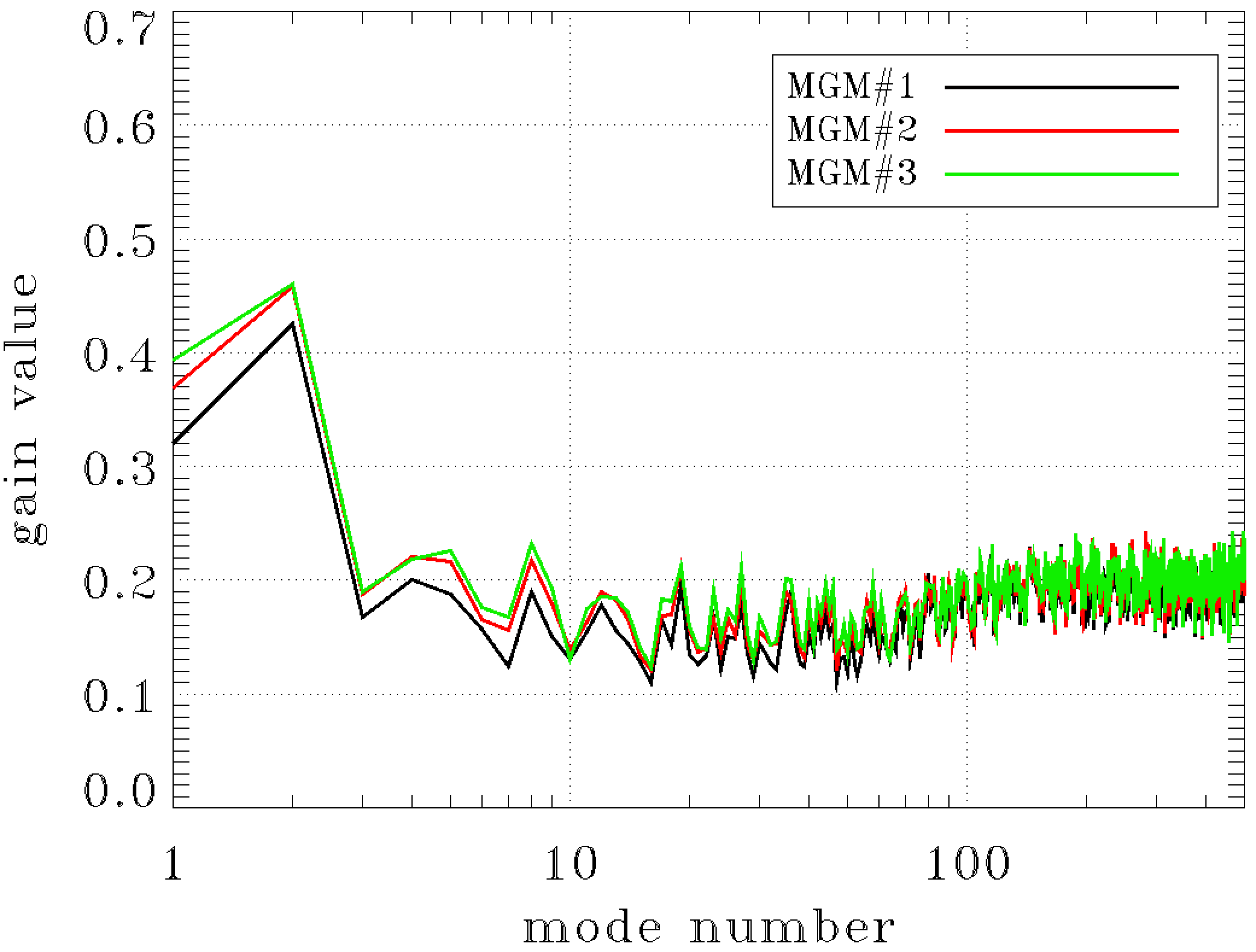}}
    \end{center}
    \caption{Examples of optimized control filter gains.
    }\label{fig:GAINdayAndNight}
\end{figure}

As for the IIR filters, MGM tests started  with the calibration source and a retro-reflector (daytime).
We injected commands on the ASM to produce a 0.6arcsec-seeing equivalent turbulence with an equivalent average wind speed of 15m/s and we tuned the calibration source to a flux equivalent to a star R11.9 and the loop frequency at 990Hz.
We ran the AO system under the same conditions with the trial-and-error gain optimization and then with MGM.
In Fig. \ref{fig:GAINday}, we report the gain vectors obtained with the two methods.
Note that with the calibration source, MGM gains are stable at the level of a few percent because the turbulence level simulated by the ASM is constant.
The exception is the Tip/Tilt gain that significantly varies following the level of vibrations (also in daytime the vibration level is varying in time due to instruments pumps/fans status and movements of the personnel in the telescope building).
MGM, optimizing mode by mode, is able to increase the gains of the modes with lower noise propagation, and in particular is able to decouple Tip and Tilt that are affected by a different level of vibrations. 
The SR of the PSFs obtained on LUCI is slightly better in the case of MGM: 60\% for the trial-and-error method an 66\% for MGM. 
The command injection on ASM, used to emulate the atmospheric disturbance, has a circular buffer of 4000 elements (1 for each loop step), this prevents the emulation of a seeing variation on timescales of tens of seconds. Then, the dynamic behaviour of MGM can be tested on sky only.  

\begin{figure*}
    \begin{center}
    \begin{tabular}{c}
        \includegraphics[width=0.66\textwidth]{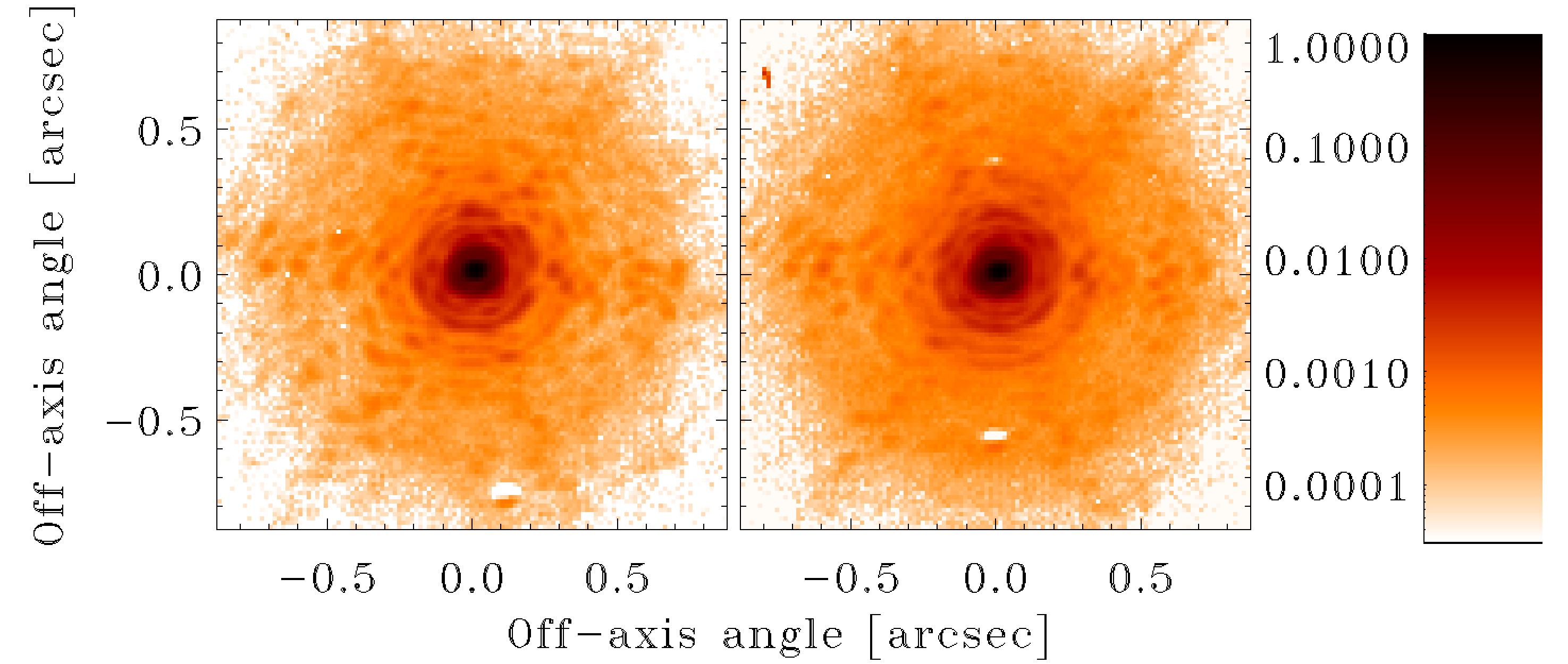}
    \end{tabular}
    \end{center}
	 \caption{\label{fig:PSFnight} LUCI FeII band PSF (40s) acquired on sky. Comparison between MGM and old optimization method on a R10.9 NGS. The loop frame rate is 1212Hz and 40$\times$40 sub-apertures are used. An average of 5ph/sa/frame are detected on the WFS valid sub-apertures. Peaks of the images are normalized to 1. Left, PSF acquired with gains optimized with MGM (SR=0.65), right, PSF acquired with gains optimized by trial-and-error method (SR=0.61). The DIMM seeing respectively is 1.07arcsec and 0.95arcsec for the two acquisitions while the average wind speed estimated from telemetry data is in the range of 12-14m/s for both acquisitions.}
\end{figure*}

We tested MGM on sky on July 2019.
We used a R 10.9 star as NGS, 
closing the loop at 1212Hz. At first we optimized the gains with the trial-and-error method, then with MGM. In both cases, we acquired long exposure (40s) PSFs at 1646nm (FeII filter) with LUCI.
Seeing during observation was quite stable around 1.0arcsec.
In Fig. \ref{fig:PSFnight} we show the PSFs and in Fig. \ref{fig:GAINnight} the gain vectors for MGM where we can see that the gains are stable in time except for the first 10-20 that are more affected by changes of vibration amplitude and seeing.
The performance of MGM is slightly better: a FeII band SR of 65$\pm$2.5\% with respect to 61$\pm$2.6\% for the old method.
The expected performance from simulations with such a seeing value is 73\%, slightly higher than the one obtained on LUCI1 PSFs, but still compatible given the uncertainty on atmospheric parameters. 

\begin{figure*}
    \begin{center}
    \begin{tabular}{c}
        \includegraphics[width=0.9\textwidth]{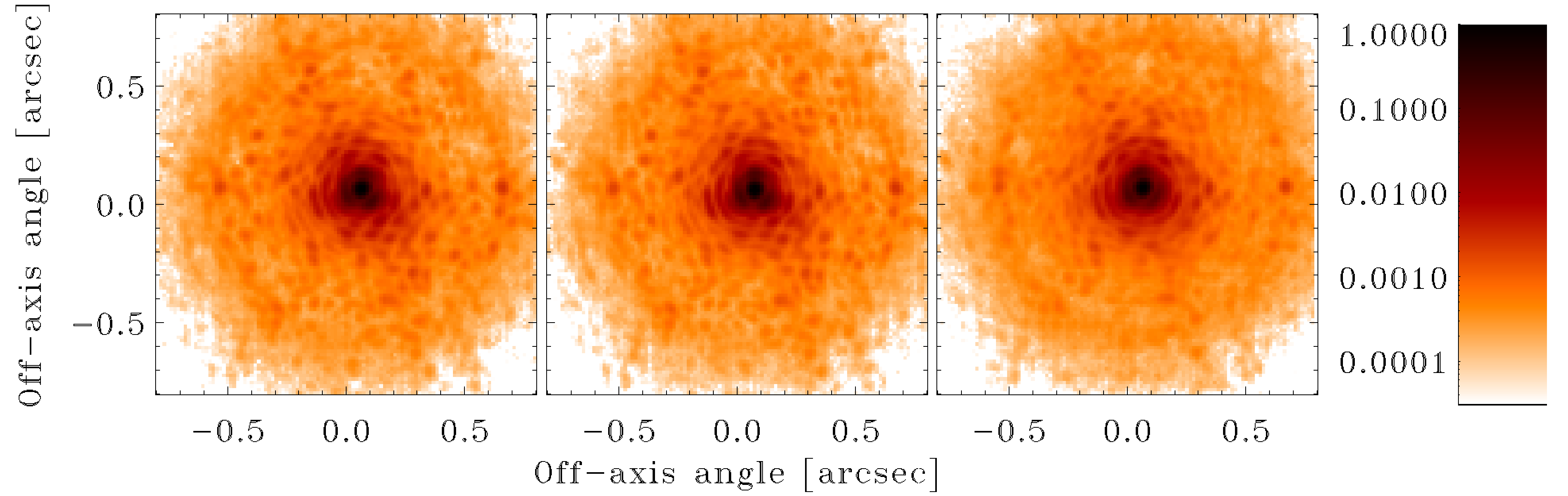}
    \end{tabular}
    \end{center}
	 \caption{\label{fig:PSFnight2} LUCI FeII band PSF (7.62s) acquired on sky. Test of MGM on a R9.4 star. The loop frame rate is 1700Hz and 40$\times$40 sub-apertures are used. An average of 25ph/sa/frame are detected on the WFS valid sub-apertures. Peaks of the images are normalized to 1. They are the one with lowest SR (0.47), the one with SR close to average (0.59) and the one with highest SR (0.64) of a series of 10 PSFs. The DIMM seeing varies from 1.16 to 1.50arcsec during the acquisition and average wind speed estimated from telemetry data from 6 to 8m/s. PSFs show a bit of non-common-path aberration from LUCI as three lobes on the first ring.}
\end{figure*}

We did several test of the algorithm in different atmospheric conditions.
For example during July 2020 we acquired several PSFs on
a R 9.4 star
with a DIMM seeing ranging between 1.2arcsec and 1.5arcsec.
In this case MGM ran together with Tip/Tilt IIR filters.
The system was stable with a SR in FeII bandwidth between 47 and 64\% (see Fig.\ref{fig:PSFnight2}, expected from simulations 48 and 67\%).

\begin{figure*}
    \begin{center}
    \begin{tabular}{c}
        \includegraphics[width=0.9\textwidth]{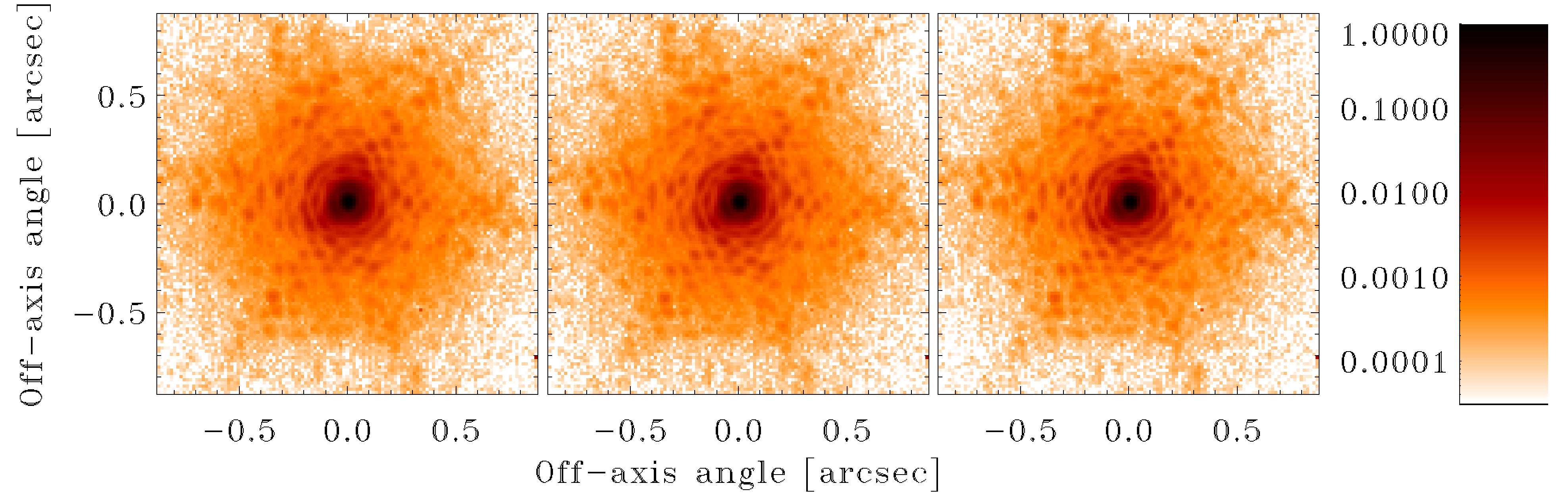}
    \end{tabular}
    \end{center}
	 \caption{\label{fig:PSFnight3} LUCI FeII band PSF (25s) acquired on sky. Test of MGM on a R13 NGS. The loop frame rate is 980Hz and 20$\times$20 sub-apertures are used. An average of 6ph/sa/frame are detected on the WFS valid sub-apertures. Peaks of the images are normalized to 1. They are the one with lowest SR (0.61), the one with SR close to average (0.64) and the one with highest SR (0.67) of a series of 9 PSFs. The DIMM seeing varies from 0.98 to 1.10arcsec during the acquisition and average wind speed estimated from telemetry data from 4 to 6m/s.}
\end{figure*}

\begin{figure*}
    \begin{center}
    \begin{tabular}{c}
        \includegraphics[width=0.66\textwidth]{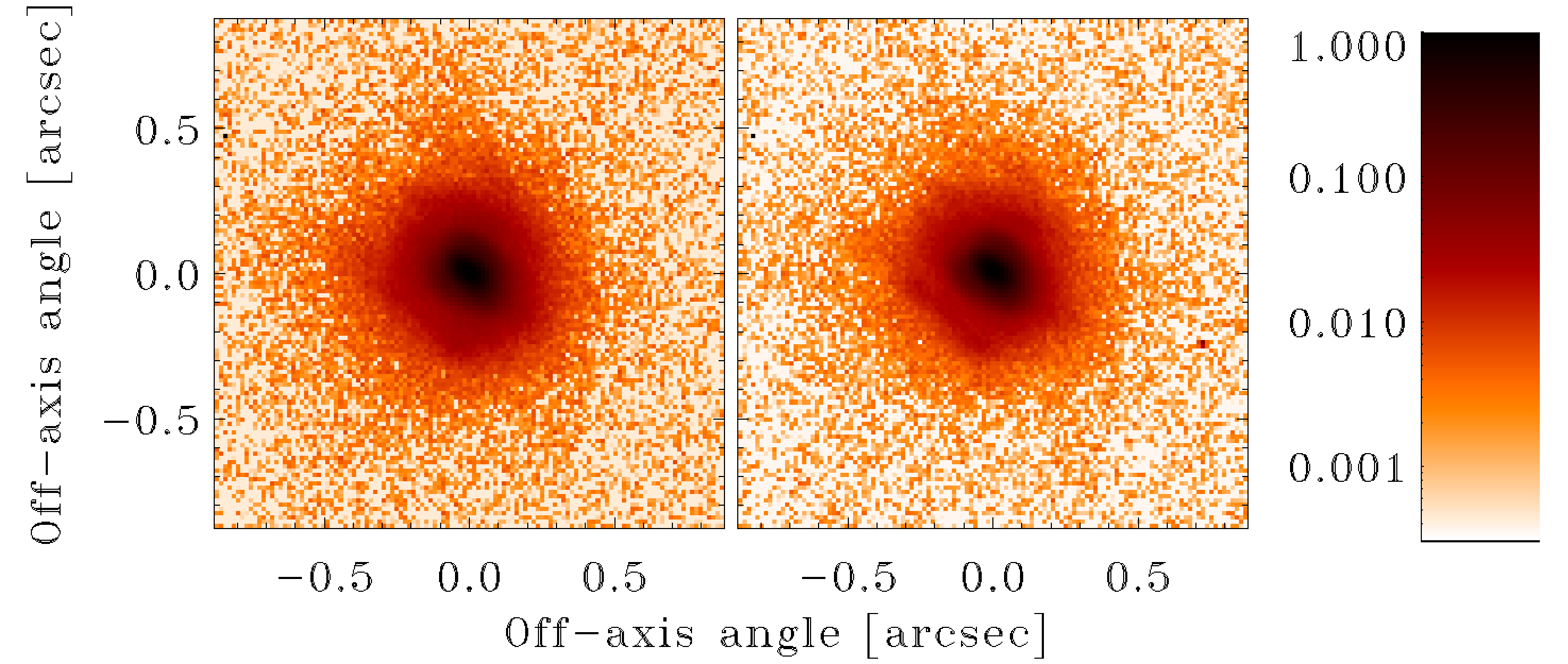}
    \end{tabular}
    \end{center}
	 \caption{\label{fig:PSFnight4} LUCI Ks band PSF (225s) acquired on sky. Test of MGM on a R16.5 NGS. The loop frame rate is 390Hz and 10$\times$10 sub-apertures are used. An average of 2ph/sa/frame are detected on the WFS valid sub-apertures. Peaks of the images are normalized to 1. The SR of the PSFs is 0.19 and 0.23. The DIMM seeing is stable at 1.00arcsec during the acquisition. Average wind speed estimated from telemetry data is not available with such dim star.}
\end{figure*}

Then, during October 2020 we tested MGM on fainter stars, on
a magnitude 13 star and on
a magnitude 16.5 star.
On these fainter stars the system is driven with the detector in higher binning mode, 20$\times$20 and 10$\times$10 sub-apertures are used instead of 40$\times$40, and with lower frame rates, 980 and 390Hz, respectively for the two magnitudes.
The DIMM seeing was quite stable around 1.0arcsec in both cases.
SR was between 61 and 67\% in FeII bandwidth (see Fig.\ref{fig:PSFnight3}, expected from simulations 58\%) for the brighter star
and between 19 and 23\% in Ks bandwidth (see Fig.\ref{fig:PSFnight4}, expected from simulations 18\%) for the fainter star.
Similarly to previous on-sky measurements, the achieved SR values are in reasonable agreement with the ones from simulation given the uncertainty on atmospheric parameters such as $\tau_0$ and L$_0$.

\begin{figure*}
    \begin{center}
    \begin{tabular}{c}
        \includegraphics[width=0.55\textwidth]{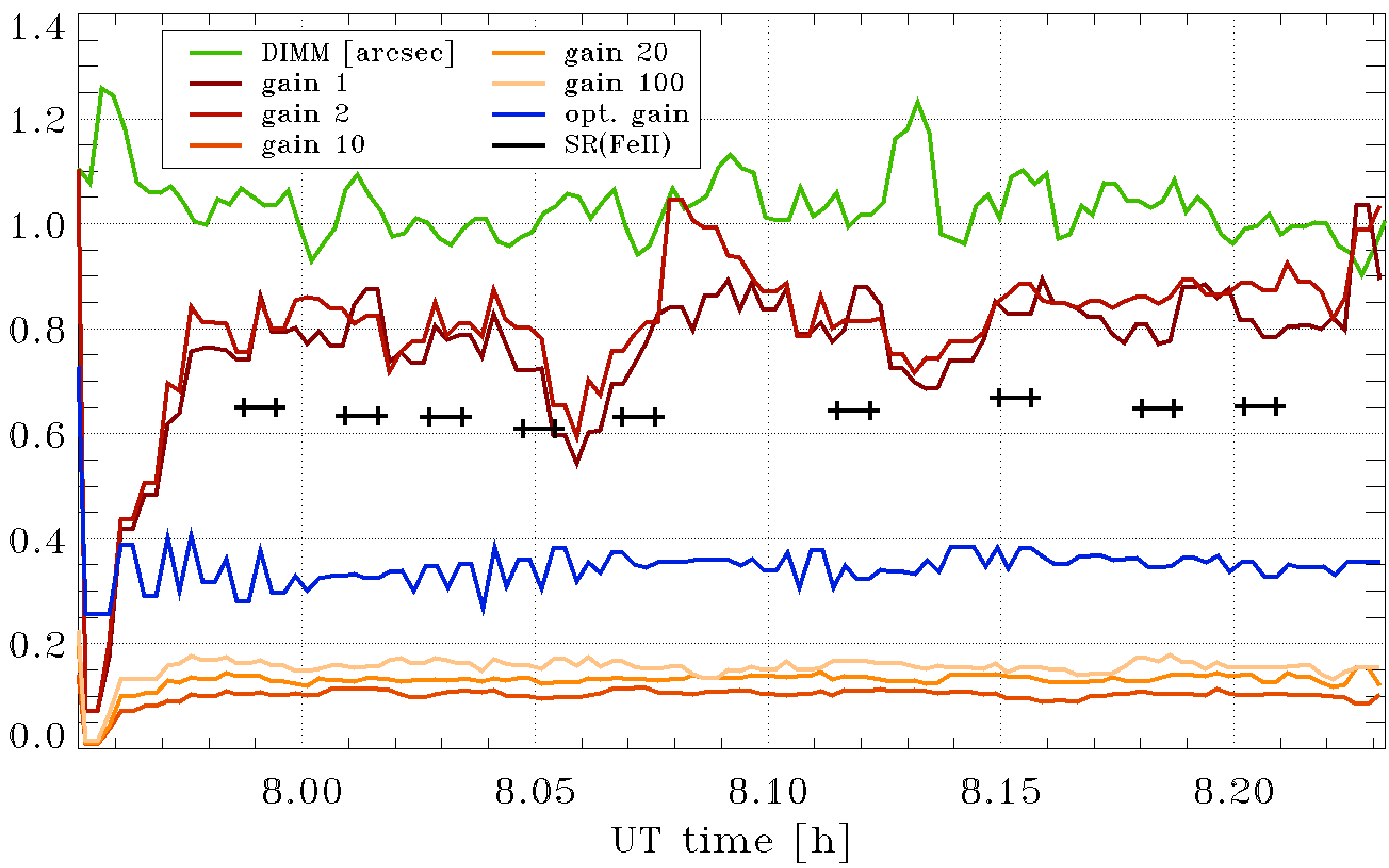}
    \end{tabular}
    \end{center}
	 \caption{\label{fig:timeSerie}
	 Time series from the second of October 2020 of DIMM seeing, a few modal gains (no. 1 is Tip), estimated optical gain and FeII bandwidth SR (from a 25seconds long integration) during a test of MGM on a R13 NGS. The loop framerate is 980Hz and 20$\times$20 sub-apertures are used. An average of 6ph/sa/frame are detected on the WFS valid sub-apertures. Fourth, sixth and seventh SRs are computed from PSFs shown in Fig.\ref{fig:PSFnight3}. Note that 5-10\% of the variability of the optical gain can be associated to its own estimation errors (the error depends on the amount of seeing and its variability), more details will be reported in a paper in preparation.}
\end{figure*}

Finally, we show in Fig.\ref{fig:timeSerie} the time series of the DIMM seeing, a few modal gains, the estimated optical gain and the FeII SR from the slow telemetry at about 1Hz acquired when the system was in closed loop with the magnitude 13 star. We can see that gains are optimized about every 5 seconds and that the transitory from initial condition with very low gains lasts about 1 minute. Note that Tip/Tilt gains (1 and 2 in Fig.\ref{fig:timeSerie}) are the most variable ones because they adapt to the continuously changing wind shake/vibration conditions of the telescope: \emph{e.g.}, the vibration at 13Hz can change in amplitude of 30-40\% in a few tens of seconds.


\section{Conclusions}\label{conclusion}

In this work we showed two advances in the control of SCAO systems: Tip/Tilt control filter to deal with structural resonances and the first on-sky demonstration of modal gain optimization with a PWFS system.

In this work we have shown how the control of SOUL is configured and optimized to reach the desired performance and take advantage of the features of its hardware, focusing mainly on two aspects: the Tip/Tilt control filters and the modal gain optimization.

Before the work described here, SOUL Tip/Tilt control was under-performing because a recently found control-structure interaction in the ASM prevented the classical integrator control to reach the expected performance limiting the gains that can be used.
Hence, we designed new temporal control filters shaping the closed loop transfer function in such a way that an effective rejection of turbulence and vibrations is provided while avoiding the excitation of this control-structure interaction.
We verified this both in daytime with a calibration source and on sky with an NGS, obtaining a gain larger than a factor two in the rejection of the main system vibration at 13Hz.
Now this feature is routinely used at the telescope during standard night time operations of the SOUL systems (both with LBTI and LUCI instruments) and it can be easily adapted to any AO system affected by similar issues.

Then, we presented our study on modal gain optimization.
We develop a general algorithm to optimize on-line the control filters gains in the case of a non-linear WFS such as the PWFS taking advantage of the previous work done on the PWFS optical gain and working in conjunction with the new Tip/Tilt control.
This algorithm, currently limited to engineering mode (but close to be ready for standard operations), proved to perform equally or slightly better than the previous trial-and-error method both on the controlled environment obtained with a calibration source and on sky during commissioning time operation.
In our knowledge this is the first time that modal gains of a SCAO system with a PWFS have been optimized on sky.
This is a key milestone for PWFS-based SCAO systems for the current 8-m class telescopes and for the future extremely large telescopes.

\section*{Acknowledgements}
The SOUL team wants to acknowledge the valuable and constant support provided by the AO group of LBTO.
We also thanks the LBTO personnel for the help during the activity on the mountain.
This work is part of the SOUL project and it is funded by LBTO under the second generation instruments program.

\section*{DATA AVAILABILITY}
The data underlying this article will be shared on reasonable request to the corresponding author.



\bibliographystyle{mnras}
\bibliography{biblio} 




\bsp	
\label{lastpage}
\end{document}